\documentclass[showpacs,preprintnumbers,amsmath,amssymb,superscriptaddress]{revtex4}

\usepackage{graphicx}
\usepackage{subfigure}
\usepackage{dcolumn}
\usepackage{bm}
\usepackage{threeparttable}
\usepackage{multirow}

\usepackage[colorlinks=true,linktocpage=true,linkcolor=blue,citecolor=blue]{hyperref}
\newcommand{\gev}{\mathrm{GeV}}
\newcommand{\mev}{\mathrm{MeV}}

\graphicspath{{Fig/}}

\begin{document}

\title{Determine the quantum numbers of $X(6900)$ from photon-photon fusion in ultra-peripheral  heavy ion collisions}

\author{Peng-Yu Niu}\email{niupy@m.scnu.edu.cn}
\address{Guangdong Provincial Key Laboratory of Nuclear Science, Institute of Quantum Matter, South China Normal University, Guangzhou 510006, China}
\affiliation{Guangdong-Hong Kong Joint Laboratory of Quantum Matter, Southern Nuclear Science Computing Center, South China Normal University, Guangzhou 510006, China}

\author{Enke Wang}\email{wangek@scnu.edu.cn}
\address{Guangdong Provincial Key Laboratory of Nuclear Science, Institute of Quantum Matter, South China Normal University, Guangzhou 510006, China}
\affiliation{Guangdong-Hong Kong Joint Laboratory of Quantum Matter, Southern Nuclear Science Computing Center, South China Normal University, Guangzhou 510006, China}

\author{Qian Wang}\email{qianwang@m.scnu.edu.cn}
\address{Guangdong Provincial Key Laboratory of Nuclear Science, Institute of Quantum Matter, South China Normal University, Guangzhou 510006, China}
\affiliation{Guangdong-Hong Kong Joint Laboratory of Quantum Matter, Southern Nuclear Science Computing Center, South China Normal University, Guangzhou 510006, China}

\author{Shuai Yang}\email{syang@scnu.edu.cn}
\address{Guangdong Provincial Key Laboratory of Nuclear Science, Institute of Quantum Matter, South China Normal University, Guangzhou 510006, China}
\affiliation{Guangdong-Hong Kong Joint Laboratory of Quantum Matter, Southern Nuclear Science Computing Center, South China Normal University, Guangzhou 510006, China}

\date{\today}

\begin{abstract}
We study the production of the $X(6900)$ through the $J/\psi J/\psi$ decay channel in the ultra-peripheral heavy ion collisions at the LHC energy region. The potential quantum numbers of $X(6900)$ would be $0^{\pm +}$ and $2^{\pm+}$. We find that the transverse momentum and the polar angle distributions of $X(6900)$ can be used to distinguish these four potential quantum numbers. These typical distributions stem from the linearly polarized photon from the fast-moving nuclei and can be measured with the LHC experiments in further. 
\end{abstract}

\maketitle

\section{Introduction}
\label{chap:introduction}
Over the past two decades, a large number of exotic states, 
which are not compatible with $q\bar q$ or $qqq$ of the traditional quark model states, 
have been discovered.
Recently, the LHCb Collaboration reported resonance-like
structures in the $J/\psi$-pair invariant mass distribution~\cite{LHCb:2018uwm,LHCb:2020bwg}
in p$+$p collisions at center-of-mass energies of $\sqrt{s}=7,~8,~13~\mathrm{TeV}$, 
corresponding to an integrated luminosity of $9~\mathrm{fb}^{-1}$.
A narrow structure around $6.9~\mathrm{GeV}$, so-called $X(6900)$, and a broad structure 
slightly above the $J/\psi J/\psi$ threshold were reported. 
Considering their mass positions and the observed channels,
the community reach a consensus that they contain
two charm and two anti-charm quarks. 
Later on, the CMS Collaboration~\cite{CMS} reported three
resonance-like structures through the same channel in p$+$p collisions at $\sqrt{s}=~13~\mathrm{TeV}$ with an integrated luminosity of $135~\mathrm{fb}^{-1}$.
The significances of the lower two structures with mass of $6927\pm 9\pm 5~\mev$ and $6552\pm10\pm 12~\mev$
are well above $5$ standard deviations. The highest structure
with a mass of $7287\pm 19\pm 5~\mev$ is with a significance of $4.1$ standard deviation. 
The first one is consistent with the $X(6900)$
reported by the LHCb Collaboration. Meanwhile,  the ATLAS Collaboration~\cite{ATLAS} also
reported similar structures. Those structures are the 
first measurements of full-charm tetraquarks, arising 
 intensive discussions among the hadron physics community. 
The $X(6900)$ can be accepted by both compact tetraquark
 ~\cite{Faustov:2021hjs,Deng:2020iqw,Chen:2022asf,Weng:2020jao,Richard:2020hdw,Sonnenschein:2020nwn,Guo:2020pvt,Berezhnoy:2011xn,Wu:2016vtq,Chen:2020lgj,Liu:2019zuc,Wang:2019rdo,Zhao:2020zjh,Chen:2020xwe,Wang:2020ols} and
 hadronic molecular pictures~\cite{Dong:2021lkh,Dong:2020nwy,Wang:2022jmb}. 
 It can be either a pseudoscalar $P$-wave state with $J^{PC}=0^{-+}$
 or a scalar $S$-wave state with $J^{PC}=0^{++}$
 ~\cite{Wu:2016vtq,Debastiani:2017msn,liu:2020eha,Wan:2020fsk,Karliner:2020dta,Lu:2020cns,Liang:2021fzr,Li:2021ygk,Ke:2021iyh}. Another possible quantum number scenario might 
 be $J^{PC}=2^{++}$ as discussed in Refs.~\cite{Faustov:2021hjs,Deng:2020iqw,Chen:2022asf,Weng:2020jao,Karliner:2020dta,Lu:2020cns,Liang:2021fzr,Li:2021ygk,Ke:2021iyh,Bedolla:2019zwg,Zhu:2020xni}.
 In the molecular picture, both $J^{PC}=0^{++}$ and $J^{PC}=2^{++}$
are allowed~\cite{Dong:2021lkh,Dong:2020nwy}. Therefore,
the determination of its quantum numbers, especially a distinguish between $0^{-+}$
and $0^{++}$ would shed new light on the nature of these recently measured full-charm tetraquark states. 

On the other hand, the Lorentz-boosted electromagnetic fields surrounding ultrarelativistic heavy ions with large charges can be treated a flux of quasireal photons\cite{Fermi:1924tc,Williams:1934ad,vonWeizsacker:1934nji}. The photon intensity is proportional to the square of the ion charge ($Z$). Exclusive photon-photon interactions, especially the $\gamma\gamma\to l^+ l^-$ processes, have been extensively studied in ultra-peripheral and hadronic heavy ion collisions~\cite{Budnev:1975poe,Bertulani:1987tz,Baur:2001jj,Bertulani:2005ru,Baur:2007zz,Baltz:2007kq,Klein:2020fmr,Zha:2018ywo,Klein:2018cjh,Azevedo:2019fyz,Klein:2020jom,Xiao:2020ddm,Brandenburg:2021lnj,STAR:2004bzo,STAR:2018ldd,STAR:2019wlg,ALICE:2013wjo,ATLAS:2018pfw,ATLAS:2020epq,CMS:2020skx}. Moreover, photon-photon interactions can be used to study the exotic particles~\cite{Goncalves:2012zm,Goncalves:2019vvo,Moreira:2016ciu,Goncalves:2018hiw,Goncalves:2021ytq,Esposito:2021ptx}, Higgs bosons~\cite{Drees:1989vq,Papageorgiu:1990mu,Zhu:2015via,dEnterria:2019jty}, and to search for physics beyond the standard model~\cite{Bruce:2018yzs,ATLAS:2017fur,ATLAS:2020hii,CMS:2018erd}. Recently, people realized that the equivalent photon has impact parameter ($b$) dependence~\cite{CMS:2020skx,Vidovic:1992ik,Zha:2018tlq,Wang:2021kxm,Klusek-Gawenda:2020eja} and is linearly polarized~\cite{Li:2019yzy,Li:2019sin}. Specifically, the linear polarized photons lead to the $\cos(4\phi)$ azimuthal asymmetry for dilepton production from photon-photon scatterings, confirmed by the STAR Collaboration~\cite{STAR:2019wlg}. This linearly polarized quasireal photons could provide an extra information about the spin of particle created by photon-photon fusion. In ultra-peripheral collisions (UPC), the nuclei pass on another with an impact parameter greater than twice the nuclear radius, which means no nuclear overlap occurs. Based on this facts, UPC provides an excellent platform to investigate the exotic states created by photon-photon fusion with the following advantages: 1) the background is significantly suppressed compared with that in hadronic collisions,  2) the cross section is enhanced by $Z^4$ compared with that in $e^+e^-$ collisions.
  
In this work, we focus on the $X(6900)$ kinematic distributions instead of absolute cross section to determine its quantum numbers, which could be $0^{++}$, $0^{-+}$, $2^{++}$ and $2^{-+}$. We demonstrate that the kinematic distributions provide a powerful tool to identify the quantum numbers of $X(6900)$ in the future measurements at the LHC. This paper is organized as follows. Sec.~\ref{chap:framework} describes the details of calculation. The results and discussions are presented in details in Sec.~\ref{chap:result}. Lastly, Sec.~\ref{chap:summary} provides a concluding summary.

\begin{figure}[!htbp]
    \centering
\includegraphics[scale=0.5]{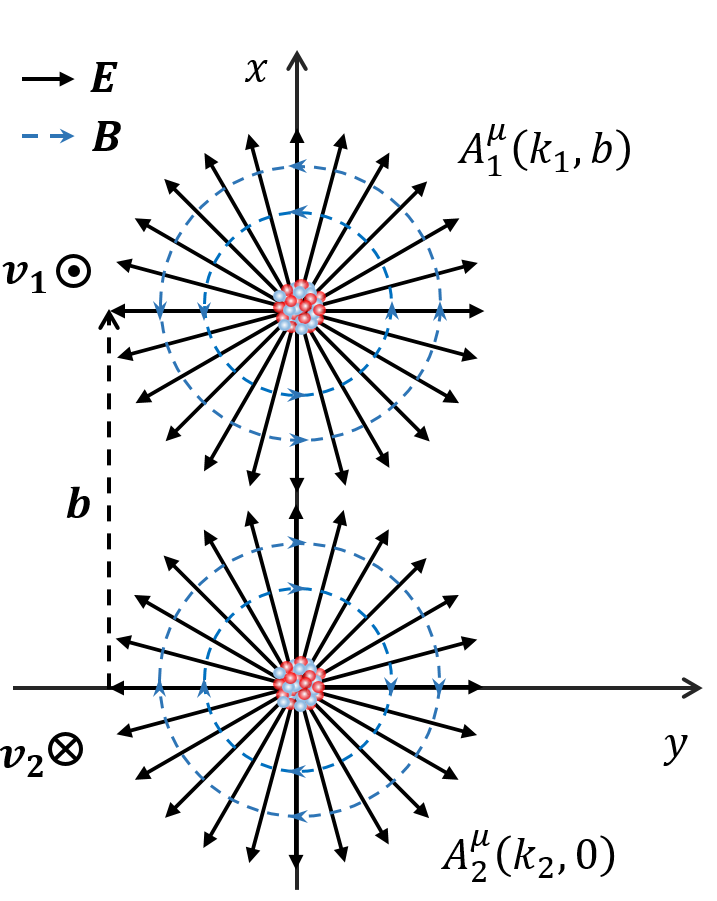}
\caption{Illustration of the electric and magnetic field lines in the plane perpendicular to the beam.
The black solid and blue dashed lines are for electric and magnetic fields, respectively.  
$v_1$ and $v_2$ are velocities of the two nuclei. $b$ is the impact parameter
defined as the distance between the two nuclei in the transverse plane with respect to the beam. }
\label{fig:EM}
\end{figure} 

\section{Framework}
\label{chap:framework}

\begin{figure}[!htbp]
    \centering
\includegraphics[scale=0.6]{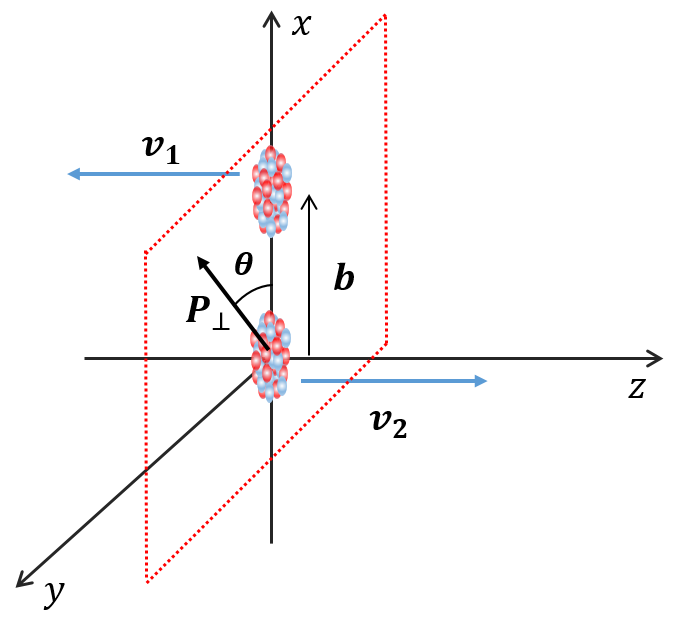}
\caption{The setup of the coordinate for the photo-photo fusion process in UPC. 
$v_1$ and $v_2$ are the velocities of two nuclei. 
The $z$-axis is chosen along beam direction. 
The direction of the nucleus-nucleus impact parameter $\vec{b}$ 
points from one nucleus to another, e.g. from nucleus with
velocity $v_2$ to nucleus with velocity $v_1$, also serving as the direction of the $x$-axis.  Thus, the $y$-axis can be determined by right hand rule. }
\label{fig:coordinate}
\end{figure}

The electromagnetic field surrounding ultra-relativistic nuclei are highly contracted and nearly perpendicular to the direction of motion, as depicted in Fig.~\ref{fig:EM}. Therefore, the photons generated by these fields are expected to be linearly polarized in the transverse plane with respect to the beam~\cite{Li:2019yzy,Li:2019sin}. The electromagnetic fields of the two nuclei are treated classically. The two nuclei are approximated to move with constant velocity on straight lines with
 an impact parameter $b$ as shown in Fig.~\ref{fig:coordinate}, with
negligible deflections due to the collision. Accordingly, 
the two electromagnetic potentials read as~\cite{Vidovic:1992ik},
 
\begin{align}
\label{eq:A}
A_{1}^\mu(k_1,\bold b_1)&=-2\pi Z_1e \delta(k_{1}\cdot u_1) \frac{F(-k_1^2)}{k_1^2}e^{i k_1\cdot b_1} \times u_1^\mu,\\
A_{2}^\mu(k_2, \bold b_2)&=-2\pi Z_2e \delta(k_{2}\cdot u_2) \frac{F(-k_2^2)}{k_2^2}e^{i k_2\cdot b_2}
\times u_2^\mu,
\end{align}
where $\delta$ function ensures that the nuclei moves along a straight line with a constant velocity. 
$k_{1}$ and $k_{2}$ are the four momenta of the two photons, respectively. 
$u_{1}=\gamma(1,0,0,-v)$ ($Z_1$) and $u_{2}=\gamma(1,0,0,+v)$ ($Z_2$) are the four velocities (nuclear charge numbers) of the two nuclei, respectively. $\gamma=(1-v^2)^{-1/2}$ is the Lorentz contraction factor.
$\bold b_{1}=(b,0,0)$ and $\bold b_{2}=(0,0,0)$ are the coordinates of the two nuclei. 
$F(-k_{1}^2)$ and $F(-k_{2}^2)$ are the charge form factors of two nuclei,
 which is the Fourier transform of its spatial charge distribution of the nucleus, as shown in Eq.~\ref{eq:ff}
 \begin{align}
 \label{eq:ff}
F(k^2)=\int d^3 \bold r \rho(\bold r) e^{i\bold k \cdot \bold r}
\end{align}
 
 Here $\bold k$ and $\bold r$ are momentum and coordinate, respectively.
 For a high-$Z$ nucleus, the spatial charge distribution in rest frame is well described by the Woods-Saxon distribution~\cite{Woods:1954zz}: 
 \begin{align}
\label{eq:woods}
\rho(r)=\frac{\rho_0}{1+\exp[(r-R_{A})/d]},
\end{align}
where $\rho_0$ is the normalization factor. $r$ is the distance between a given point and the nucleus center. $d$ and $R_{\mathrm{A}}$
are the skin depth and the radius of a nucleus, respectively. 
The employed values of those two parameters in this work are summarized in Tab.~\ref{table:b_RWS}.
However, it is hard to perform the Fourier transform analytically to obtain $F(k^2)$.
Alternatively, an approximation 
\begin{align}
F(|k|)=\frac{4\pi \rho^0}{A |k^3|} \left(\frac{\sin(|k|R_A)-|k|R_A\cos(|k|R_A)}{1+a^2 k^2}\right)
\end{align}
is widely employed in literatures~\cite{Klein:1999qj}. 
Here $A$ is the mass number of nucleus and $R_A=1.1 A^{1/3}~\mathrm{fm}$ 
is the radius of Woods-Saxon distribution with a hard sphere~\cite{Li:2019yzy} ($R_A=1.2A^{1/3}~\mathrm{fm}$ is also used in Refs.~\cite{Baur:1998ay,Zhu:2015via}).
 $a$ is the Yukawa potential range which is characterized by the inverse of the pion mass,
  i.e. $0.7~\mathrm{fm}$~\cite{Davies:1976zzb}. $\rho_0$ is also considered
  as a normalization factor, ensuring $F(0)=1$.

\begin{table}
   \caption{The employed values of the skin depth $d$ and the radius $R_{\mathrm{WS}}$ of Au and Pb. 
}
  \begin{tabular}{|c|c|c|c|c|}
\hline 
\multirow{2}{*}{} & \multicolumn{2}{c|}{Ref.~\cite{Shou:2014eya}} & \multicolumn{2}{c|}{Ref.~\cite{Zha:2018tlq,Li:2019yzy}}\tabularnewline
\cline{2-5} \cline{3-5} \cline{4-5} \cline{5-5} 
 & Au & Pb & Au & Pb\tabularnewline
\hline 
\hline 
$d~[\mathrm{fm}]$ & 0.41 & 0.45 & 0.535 & 0.546\tabularnewline
\hline 
$R_{\mathrm{A}}~[\mathrm{fm}]$ & 6.42 & 6.66 & 6.38 & 6.62\tabularnewline
\hline 
\end{tabular}
\label{table:b_RWS}
\end{table}

\begin{figure}[!htbp]
    \centering
\includegraphics[scale=0.5]{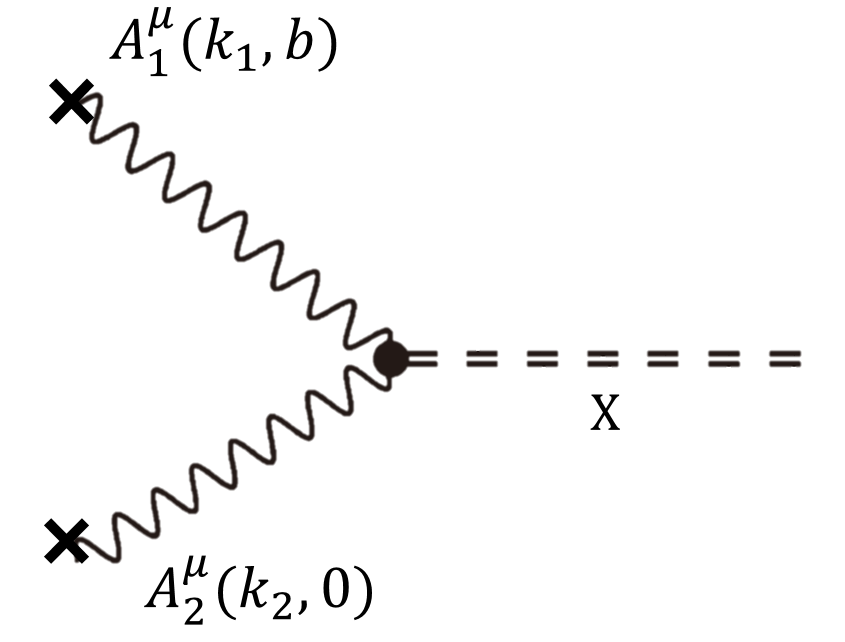}
\caption{A Feynman-like diagram for the production of the $X$  with two classical external fields $A_{1,2}^\mu$ }
\label{fig:OMP}
\end{figure}

A Feynman-like diagram for the production of the $X$ in photon-photon fusion process
is shown in Fig.~~\ref{fig:OMP}. For photon-photon fusion process, the nuclei can be assumed to move along
straight lines and their deflection can be neglected, as the transverse momentum is much smaller than
the longitudinal momentum. Therefore, the emitted photon fields can be treated as EM external fields, and the production amplitude can be expressed as\cite{Vidovic:1992ik}
\begin{align}
\label{eq:M}
\mathcal M(k_1,k_2,P)
&=\int \frac{d^4 k_1}{(2\pi)^4}\frac{d^4 k_2}{(2\pi)^4} \left[A_1^\mu(k_1,b)\Gamma_{\mu\nu}(k_1,k_2)A_2^\nu(k_2,0)\right] (2\pi)^4 \delta^4(k_1+k_2 -P)
\notag \\
&=Z^2 \frac{e^2}{\gamma^2} \int \frac{d^4 k_1}{(2\pi)^2} \delta(k_1^0-vk_1^3)\delta(P^0+vP^3-k_1^0-vk_1^3)  \frac{F(-k_1^2)}{-k_1^2}\frac{F(-k_2^2)}{-k_2^2} u_1^\mu u_2^\nu\Gamma_{\mu\nu}(k_1,k_2)e^{-i\bold b\cdot \bold k_{1,\perp}}
\notag \\
&=Z^2\frac{e^2}{2v\gamma^2}\int \frac{d^2 \bold k_{1,\perp}}{(2\pi)^2}\frac{F(-k_1^2)}{-k_1^2}\frac{F(-k_2^2)}{-k_2^2}
u_1^\mu u_2^\nu \Gamma_{\mu\nu}(k_1,k_2)e^{-i\bold b\cdot \bold k_{1,\perp}} |_{k_2=P-k_1},
\end{align}
where $P=(P^0,\bold P_\perp,P^3)$ is the four momentum of the produced $X$ particle. Here the subscript ``$\perp$" is used to label the transverse component and $\bold P_\perp=(P^1,P^2)$. 
$\delta^4(k_1+k_2-P)$ is the requirement of the four momentum conservation. $\Gamma_{\mu\nu}$ is the vertex function which depends on the quantum number of created particles and will be discussed in details later. The two photon momenta $k_1$ and $k_2$ read as
  \begin{align}
k_1=(\omega_1,\bold k_\perp, \omega_1/v),\quad k_2=(\omega_2,\bold P_\perp-\bold k_\perp,-\omega_2/v),
\end{align}
with $\omega_{1}=1/2(P_0+ v P_z)$ and $\omega_{2}=1/2(P_0- v P_z)$ the energies of the 1st and 2nd photons, respectively. Because the transverse momentum is much smaller than the longitudinal momentum, the two photons can be approximately on-shell.
$\omega_{1,2}$ and $\bold k_\perp$ is the integration variable.
Considering that $|\bold k_\perp|_{\text{min}} =0$, $k_1$ and $k_2$ are space-like vectors,
for instance
\begin{align}
k_1^2= \omega_1^2-\bold k_{1 \perp}^2 -\omega_1^2 /v^2
\le \omega_1^2-\omega_1^2 /v^2
\approx -\left(\frac{\omega_1}{\gamma} \right)^2.
\label{eq:k2}
\end{align}

In $e^+e^-$ collisions, the deflection of the scattering electron or positron
leads to finite virtuality of the emitted photon. In contrary, the heavy ions
maintain their straight-line trajectory in UPC. Thus, the low-energy
photon is regulated by the finite Lorentz contraction factor $\gamma$ of the ions,
i.e. $|k^2|\ge (\omega/\gamma)^2$ with $\omega$ the photon energy,
avoiding singularity in Eq.~\eqref{eq:M}. 
The high-energy photon is naturally cut off by the finite size of the ion's charge distribution,
i.e. $|k^2|\le (1/R_A)^2$, in which $R_A$ is the nuclear radius~\cite{Vidovic:1992ik}. Those are consistent with Eq.~\eqref{eq:k2}.
 For a given nucleus, the transverse momentum of photon is constrained by $k_\perp^2\le 1/R_A^2-(\omega/\gamma)^2$.

In this paper, we consider the production of hadrons with all the possible quantum numbers,
i.e. $0^{\pm+}$ and $2^{\pm+}$, in photon-photon fusion process.
 Hadrons with spin-1 cannot be produced due to Landau-Yang theorem~\cite{Yang:1950rg}.
 For scalars~\cite{Vidovic:1992ik} and pseudoscalars, the vertex functions are
 \begin{align}
\Gamma^{0^{++}}_{\mu\nu}=g_s(k_1 \cdot k_2 g_{\mu\nu}-k_{1\nu}k_{2\mu}),
\end{align}
\begin{align}
\Gamma^{0^{-+}}_{\mu\nu}=g_{sp}\epsilon_{\mu\nu\alpha\beta} k_1^\alpha k_2^\beta,
\end{align}
respectively. $k_1$ and $k_2$ are four momenta of the two photons. 
$g_s$ and $g_{sp}$ are the corresponding effective couplings. 
The vertex functions for $2^{\pm +}$ hadrons are
\begin{align}
\Gamma^{2^{++}}_{\mu\nu}=g_t \left(g_{\mu\nu} k_{1\rho} k_{2\sigma} \epsilon^{\rho\sigma}+
k_{1}\cdot k_{2} \epsilon^{\mu\nu}- k_{1\mu} k_{2\rho}\epsilon^{\rho\nu}- k_{2\mu} k_{1\rho}\epsilon^{\rho\nu}
\right) ,
\end{align}
and
\begin{align}
\Gamma^{2^{-+}}_{\mu\nu}=g_{tp}\epsilon_{\mu\lambda\alpha\beta} k_1^\alpha k_2^\beta\epsilon^{\lambda}_\nu.
\end{align}
  The polarized vector of spin-2 particle reads as
 \begin{eqnarray}
 \epsilon_{\mu\nu}(\vec{p},\lambda)=\sum_{\lambda_1,\lambda_2}\langle 1\lambda_1; 1\lambda_2|2\lambda \rangle \epsilon_\mu(\vec{p},\lambda_1) \epsilon_\nu (\vec{p},\lambda_2)
 \end{eqnarray}
 with $\lambda$ indicating the third component of spin. 
 $\vec{p}$ is the three momentum of the produced tensor. If the vertex function $\Gamma_{\mu\nu}$ contracts with two real photons, it is easy to verify that the amplitudes satisfy gauge invariance. 
With all the amplitude ready, one can obtain the 
differential cross section\cite{Vidovic:1992ik}  
\begin{align}
d \sigma= d^2 \bold b |\mathcal M(k_1,k_2,P)|^2 \frac{d^4 P}{(2\pi)^4} d\alpha,
\end{align}
where $d\alpha$ is the n-body phase space
\begin{align}
d\alpha=\delta^4(P-\sum_{i=1}^n p_i)\frac{d^3 \bold p_1}{(2\pi)^3 2 E_1}\frac{d^3 \bold p_2}{(2\pi)^3 2 E_2}\cdots \frac{d^3 \bold p_n}{(2\pi)^3 2 E_n}.
\end{align}
In this paper, only one-body phase space should be considered. Thus, the differential cross section can be simplified to the following form:
\begin{align}
\sigma&=\frac{2}{(2\pi)^3 2|P_z|} \int d^2 \bold b |\mathcal M(k_1,k_2,P)|^2 d P_0 d^2 \bold P_\perp \theta(E_0)|_{P_z=\sqrt{P_0^2-\bold P_\perp^2-M_X^2}},
\end{align}
where $M_X$ is the mass of hadron $X$. 
As the effective couplings $g_s,g_{sp},g_{t}$ and $g_{tp}$ are not determined yet,
 the total cross section cannot be accurately predicted. The cross section of the $X(6900)$
 in photon-photon fusion process has been estimated by Vector Meson Dominance model
 in Ref.~\cite{Esposito:2021ptx}.  This work focuses on the transverse momentum $P_\perp$
 distribution $d\sigma/(d P_\perp)$ and the polar angle distribution $d\sigma/(d \theta)$ of $X(6900)$,
 which can be used to distinguish various quantum numbers. Those two distributions
 can be obtained by integrating out another dimension from 
\begin{align}
\frac{d\sigma}{\bold d|P_\perp| ~d\theta} \propto \int d^2 \bold b d P_0 
|\bold P_\perp||\mathcal M(k_1,k_2,P)|^2 |_{P_z=\sqrt{P_0^2-\bold P_\perp^2-M_X^2}} ,
\label{eq:differential_cross_section}
\end{align}
where $\theta$ is the azimuth angle of $\bold P_\perp$ as illustrated by Fig.~\ref{fig:coordinate}.
Because the value of total transverse momentum $\bold P_\perp$ is small in UPC, $P_z$ is treated as a constant number in the last step.

\section{Results and Discussions}
\label{chap:result}
The collision time of UPC is about $t_{\text{coll}}=R_A/\gamma$, and the maximum energy of the photon is about $\omega_\mathrm{max}\approx \gamma/R_A$\cite{Vidovic:1992ik} due to the uncertainty principle. 
Taking the Pb beam with the energy per nucleon $\sqrt{s_{NN}}=5.628~\mathrm{TeV}$ 
as an example, the maximum photon energy can reach $\sim 90~\gev$ with 
$\gamma=3000$ and $R_A=6~\mathrm{fm}$~\cite{Brandenburg:2021lnj}. Such a large energy can 
 produce hadron with a mass as large as the $X(6900)$
 discussed in Sec.~\ref{chap:introduction}.  The production of the
 $X(6900)$ has been studied in heavy ion collision~\cite{Goncalves:2021ytq,Esposito:2021ptx}.
 In principle,  the photon energy should be integrated out to estimate
 the cross section of the $X(6900)$. However, we have checked that the transverse momentum $P_\perp$ and
  the polar angle $\theta$ distributions,
   which essentially help to pin down the quantum numbers of the interested hadrons,
  is not sensitive to the two-photon center-of-mass energy $\sqrt{s_{\gamma\gamma}}$ 
    in UPC (hereafter $\sqrt{s_{\gamma\gamma}}$ will be abbreviated as $\sqrt{s}$). 
    So, in this work, $\sqrt{s}$ is set to $20~\mathrm{GeV}$ as an illustration. 
    The lower limit and upper limit of the impact parameter $b=|\bold b|$ are
 twice nuclei radius and $+\infty$, respectively. Since the cross section 
 decreases dramatically with the increasing $b$~\cite{Vidovic:1992ik}, 
 the upper limit of the impact parameter $b$ is set to be $200~\mathrm{fm}$. 
 We have checked that when this value 
 increases to be $300~\mathrm{fm}$ and $400~\mathrm{fm}$, 
 our results do not change. In addition, the upper limit of 
 $|\bold k_{1\perp}|$ is set to $0.05~\mathrm{GeV}$
  which is larger than the value used in Refs.~\cite{Vidovic:1992ik,Baur:1998ay,Brandenburg:2021lnj}.
With all the parameter constrains imposed, we can 
obtain the $P_\perp$ and $\theta$ distributions from Eq.~\eqref{eq:differential_cross_section} 
for all the possible quantum numbers, i.e. $J^{PC}=0^{\pm +}$ and $J^{PC}=2^{\pm +}$.

\begin{figure}[!htbp]
\centering
\includegraphics[scale=0.5]{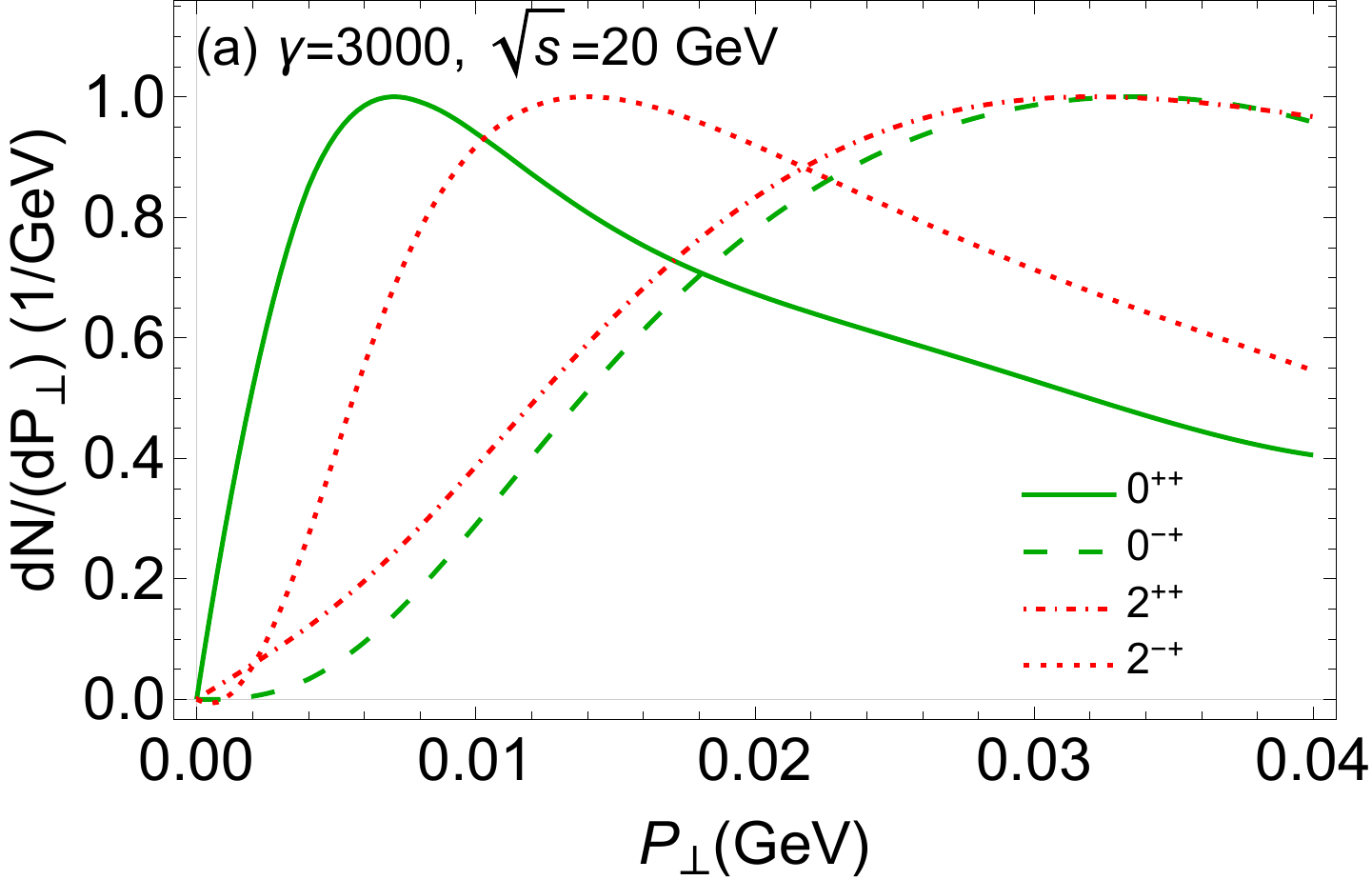} \includegraphics[scale=0.5]{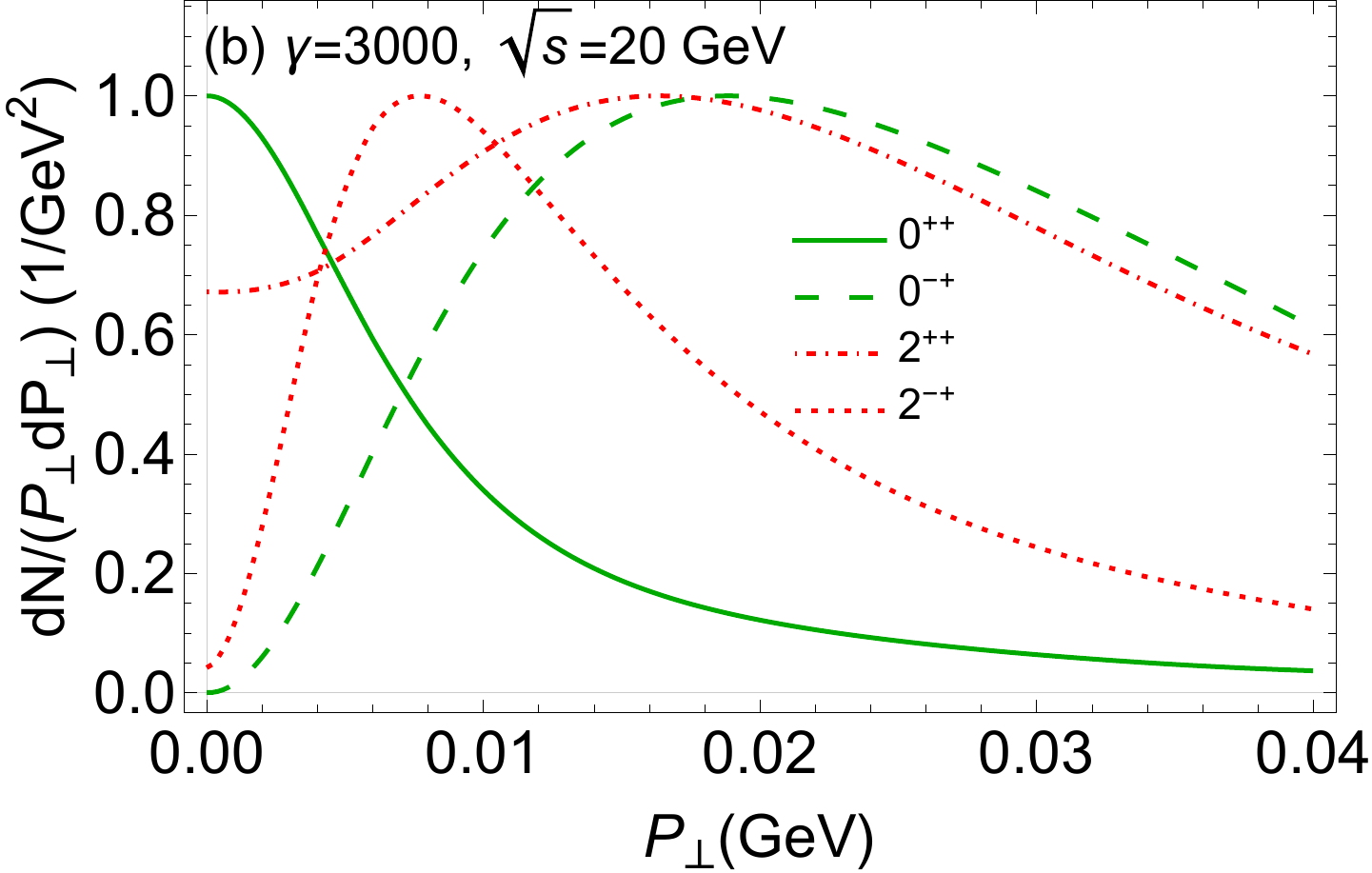}
\caption{The differential $\frac{\mathrm{d}\sigma}{\mathrm{d}P_\perp}$ (panel a)
and $\frac{\mathrm{d}\sigma}{P_\perp \mathrm{d}P_\perp}$ (panel b) distributions with polar angle $\theta$ and the impact parameter $b$ integrated out.
As all the effective couplings $g_s$, $g_{sp}$, $g_{t}$, $g_{tp}$ are not determined,
all the curves are rescaled to make the largest value to be one.
The green solid and green dashed curves are for the $0^{++}$ and $0^{-+}$ states, 
respectively. The red dot-dashed and the red dotted curves are for the $2^{++}$ and $2^{-+}$ states, respectively.}
\label{fig:Pt}
\end{figure}

The differential $\frac{\mathrm{d}\sigma}{\mathrm{d}P_\perp}$
and $\frac{\mathrm{d}\sigma}{P_\perp \mathrm{d}P_\perp}$ distributions with polar angle $\theta$ and the impact parameter $b$ integrated out, are presented in Fig.~\ref{fig:Pt}. In Fig.~\ref{fig:Pt}(b), the states with negative
parity goes to zero faster than those with positive parity around $P_\perp=0$. 
That is because the anti-symmetric tensor in the vertex function for the states with negative parity 
contracts with the photon momenta and the polarization vector/tensor of a given state. 
When  $\bold P_\perp=0$,  $k_{1\perp}$ and $k_{2\perp}$ 
are symmetric, i.e. $\bold k_{1 \perp}=-\bold k_{2 \perp}$, and the amplitudes of 
the  $0^-$ and $2^-$ states obviously go to zero. That is the reason why the 
differential $P_{\perp}$ distributions go to zero faster than those with positive parity. 
This can be used to distinguish the states with different parities. 
One can also see in  Fig.~\ref{fig:Pt} that the 
differential $P_{\perp}$ distributions of the same parity states
 have similar behavior because of the similar structure of the vertex function.
  While the four curves also share the same characteristic 
that the differential cross section decreases with the large increasing $ P_\perp$.

\begin{figure}[!htbp]
\centering
\includegraphics[scale=0.6]{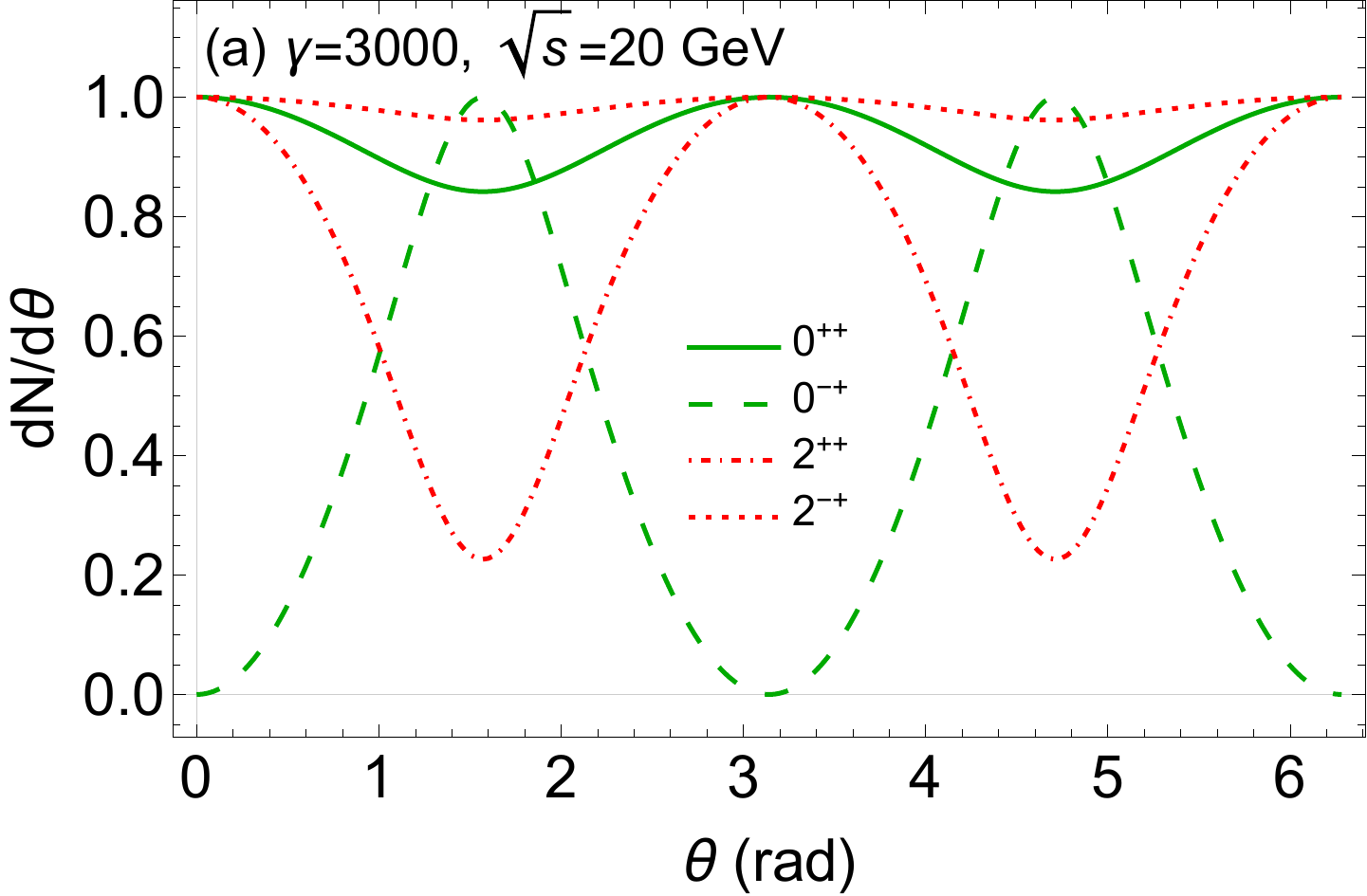} 
\caption{The differential polar angle $\theta$ distributions with the transverse momentum $\bold P_\perp$ and the impact parameter $b$ integrated out. All the curves are also rescaled to make their large values to be one. The labels of the four curves have the same meaning as those in Fig.~\ref{fig:Pt}.}
\label{fig:theta}
\end{figure}

We borrow the idea for the oscillation behavior of the dilepton pair distribution in UPC
~\cite{Li:2019yzy,Li:2019sin,Klein:2016yzr} and also present 
the differential polar angle $\theta$ distributions in Fig.~\ref{fig:theta}. They oscillate with the period $\pi$ for all the cases.
  There is no doubt that the system is symmetric with respect to
   the plane the two heavy ions locate at, 
    i.e. the $x-z$ plane in Fig.~\ref{fig:coordinate}.
    The $0^+$, $2^+$, $2^-$ distributions are ``W"-type and
    the $0^-$ distribution is ``M"-type. This can be used to distinguish
    the $0^{-}$ quantum number from the others. The oscillation strengths
    of the $0^-$ and $2^+$ states are larger than those of the other two scenarios.

\begin{figure}[!htbp]
    \centering
\includegraphics[scale=0.5]{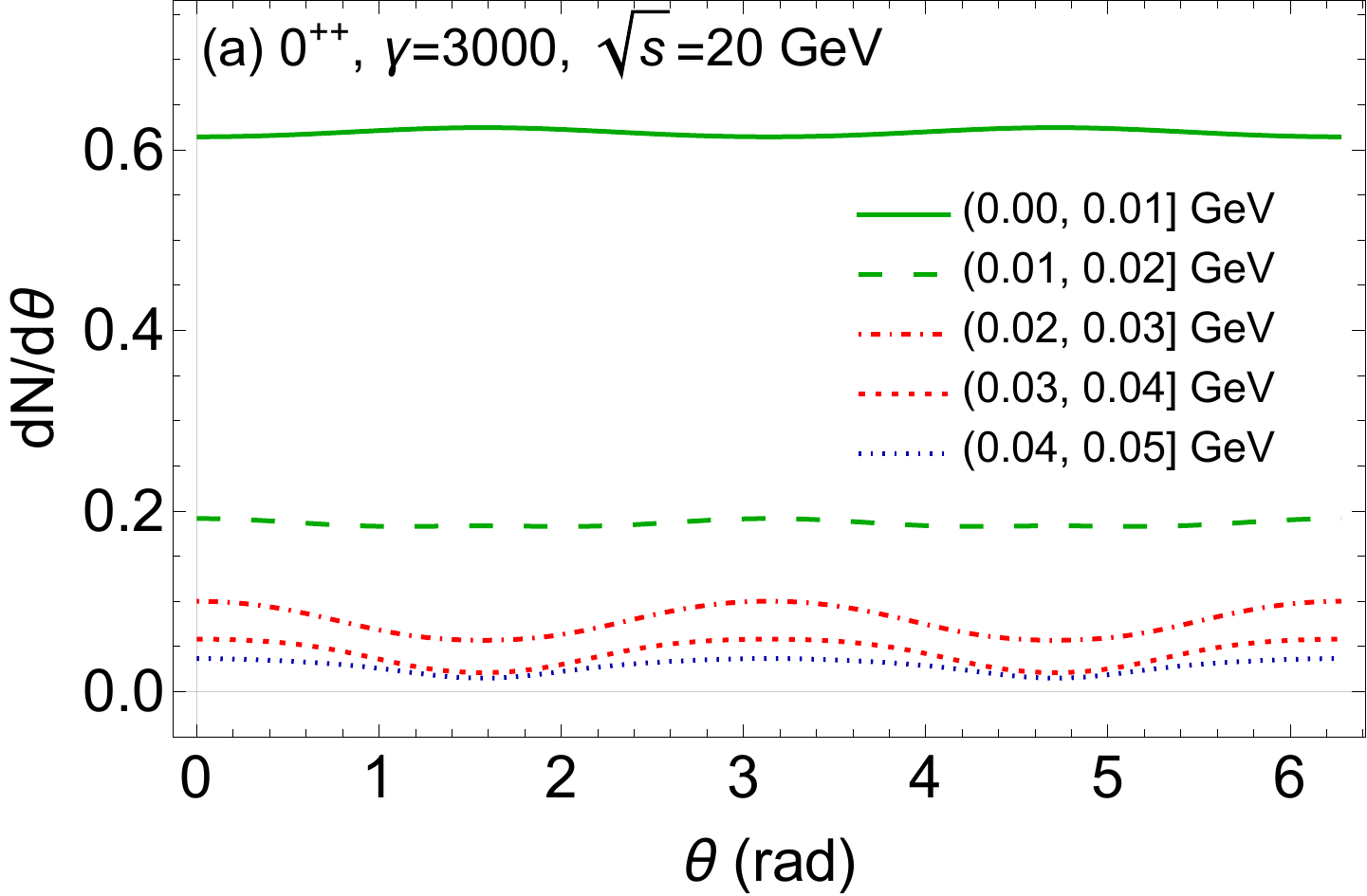}\includegraphics[scale=0.5]{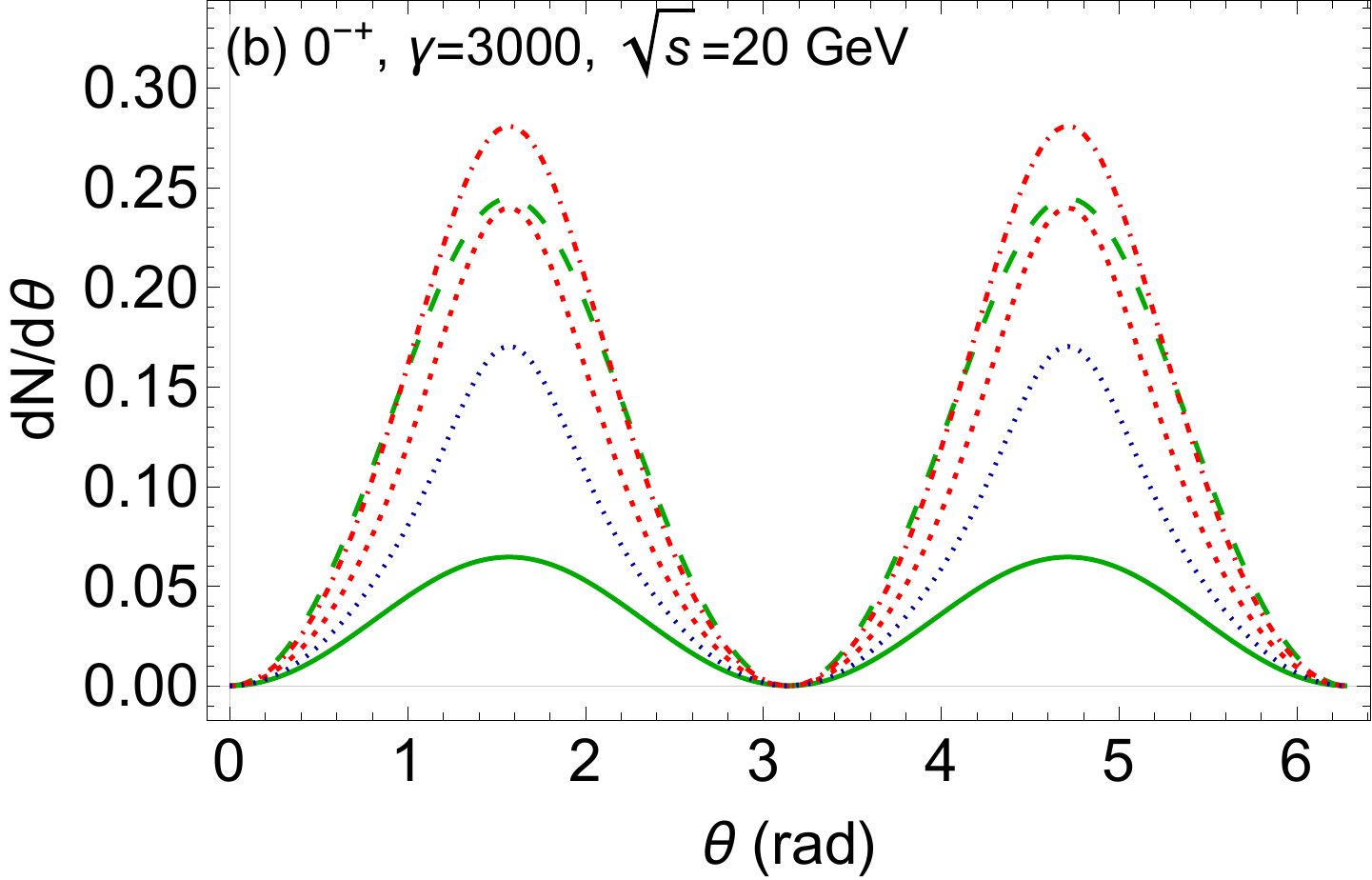}\\
\includegraphics[scale=0.5]{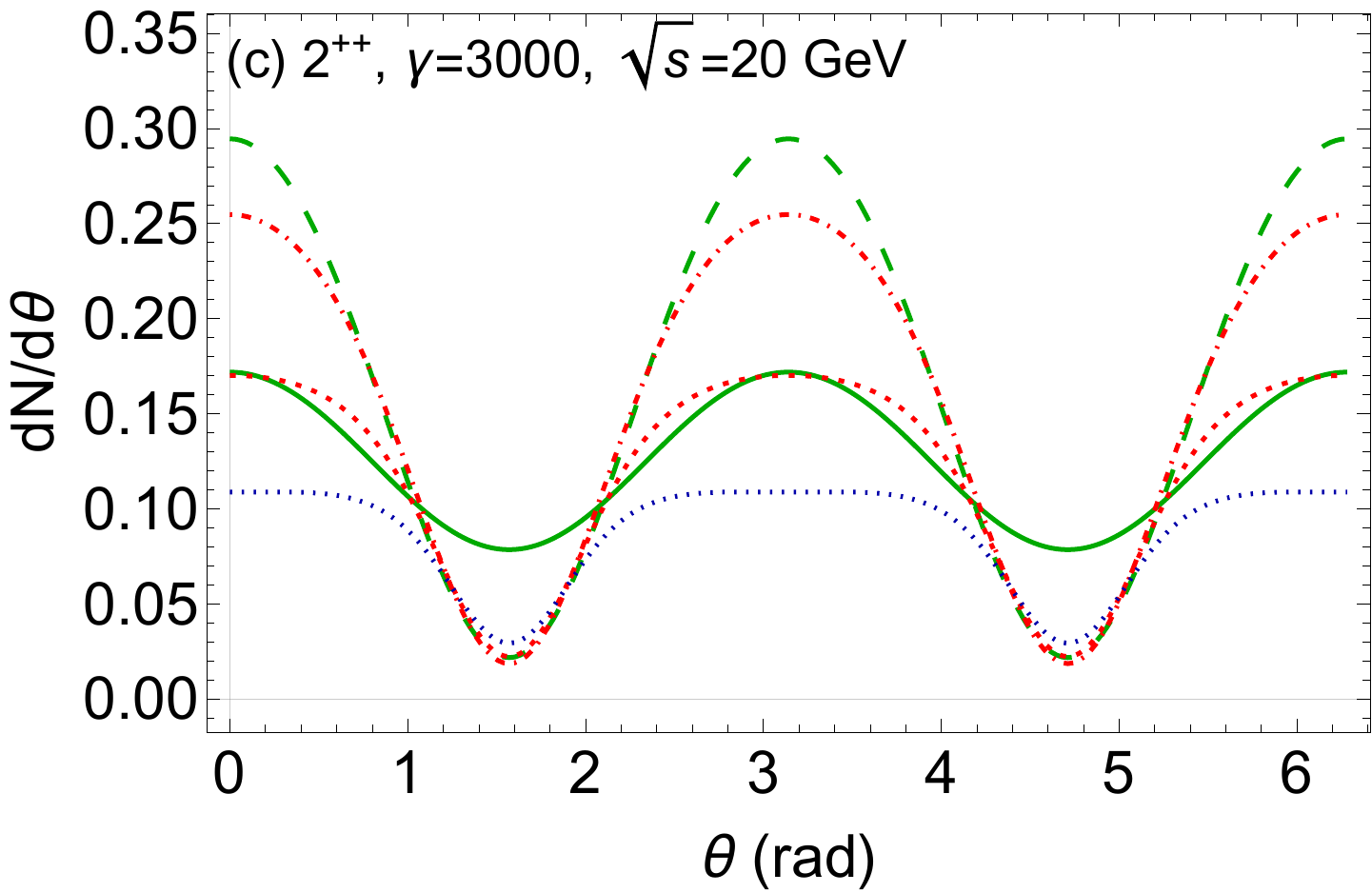}\includegraphics[scale=0.5]{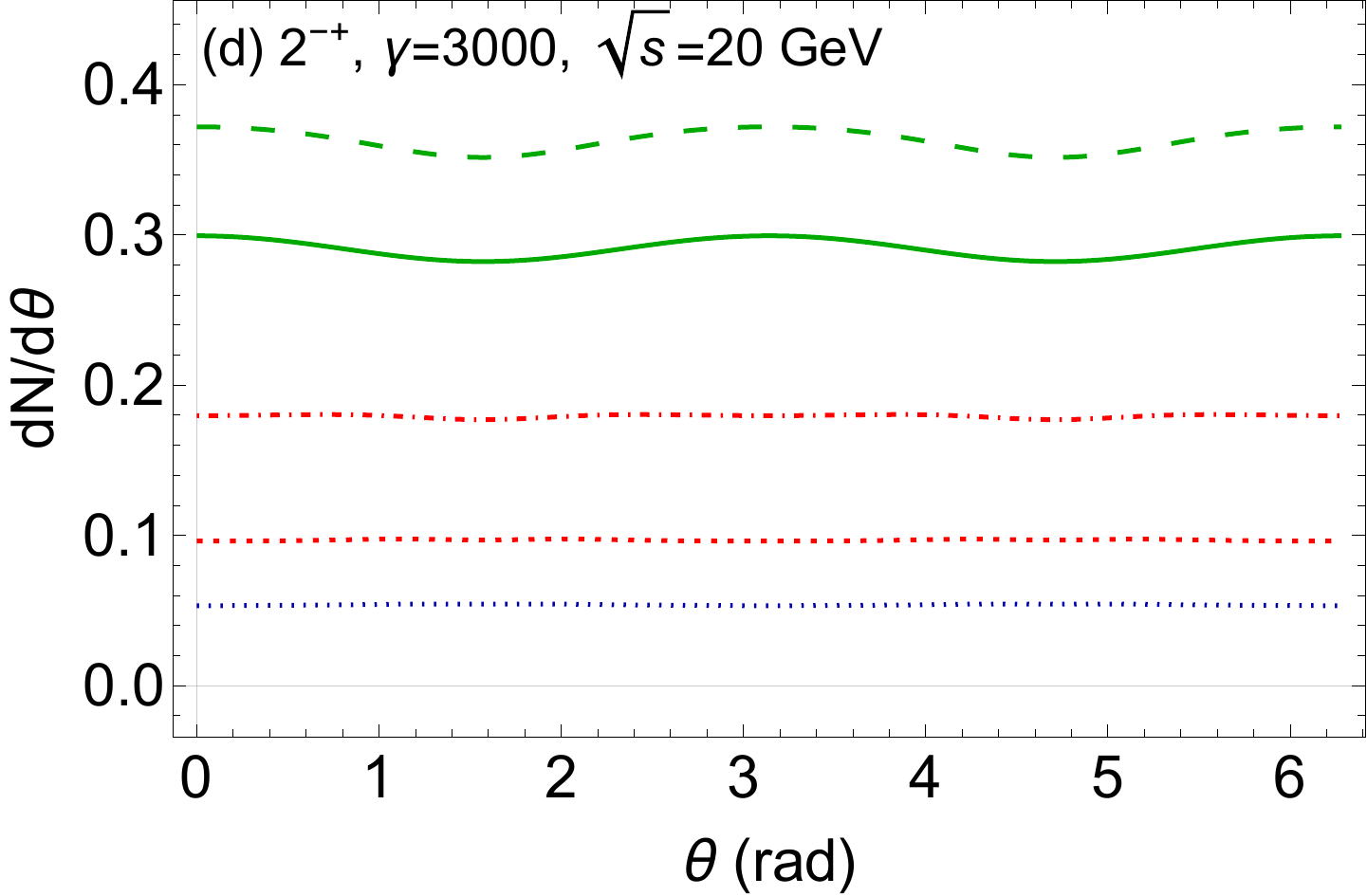}
\caption{The differential polar angle $\theta$ distributions for various transverse momentum regions,
 with the Lorentz contraction parameters $\gamma=3000$ and the two-photon center-of-mass energy 
$\sqrt s=20~\mathrm{GeV}$. The green solid, green long dashed, red dot-dashed,
 red dashed and blue dotted curves are used to label the distributions of the transverse momentum regions $(0.00, 0.01], $(0.01, 0.02], $(0.02, 0.03], $(0.03, 0.04], and $(0.04, 0.05]~\gev$, respectively. }
\label{fig:theta1q2}
\end{figure}

Fig.~\ref{fig:theta1q2} shows the differential  polar angle $\theta$ distributions for the transverse momentum regions 
 $(0.00, 0.01], $(0.01, 0.02], $(0.02, 0.03], $(0.03, 0.04], and $(0.04, 0.05]~\gev$.
 The $0^{-+}$ and $2^{++}$ states oscillate more significant than the $0^{++}$ and $2^{-+}$ states,
 which has already be seen from Fig.~\ref{fig:theta}.  The $0^{-+}$ distributions for all the $P_\perp$
 regions are ``M" type, and the distributions for other states for all the $P_\perp$
 regions are ```W" type. The oscillation for the $2^{++}$ state is more significant 
 than those of the $0^{++}$ and $2^{-+}$ states. The largest values of the $0^{++}$ and $2^{-+}$
 states are the $(0.00, 0.01]~\gev$ and $(0.01, 0.02]~\gev$ regions, respectively.
 This behavior can be used to distinguish these two states. 
 
 The above distributions, which could be measured at the LHC, can be used to determine the quantum numbers of the $X(6900)$. As our framework
  does not depend on the property of the $X(6900)$, it cannot help to shed light on its
  internal structure, for instance compact tetraquark or hadronic molecule. On the other hand,
  our framework does not only work for the $X(6900)$, but also works for other regular and exotic hadrons. 

\section{Summary}
\label{chap:summary}
Since the observation of the first full-charm tetraquark
 $X(6900)$, the community intensively discuss its internal structure,
i.e. either compact tetraquark or hadronic molecule. For the compact tetraquark scenario, all the $J^{PC}=0^{\pm+}$ and $2^{++}$ quantum numbers are allowed.  However, 
its quantum number cannot be well established due to the low statistic in current experiments. We take advantage of the polarized photon
produced in UPC and predict the transverse momentum and polar angle distributions of the $X(6900)$.
 These two distributions exhibit significant difference for the four potential quantum numbers $J^{PC}=0^{\pm+},~2^{\pm+}$,
 which can be used help to determine the quantum number of the $X(6900)$. 
 As this method does not depend on the internal structure of the interested hadrons,
 it can also work for other hadrons with spin other than one.  
  
\begin{acknowledgments}
The discussion with Hui Zhang is acknowledged. 
This work is partly supported by Guangdong Major Project of Basic and Applied Basic Research No.~2020B0301030008,
the National Natural Science Foundation of China with Grant ~No.~12147128, No.~12035007, Guangdong Provincial funding with Grant No.~2019QN01X172.
Q.W. is also supported by the NSFC and the Deutsche Forschungsgemeinschaft (DFG, German
Research Foundation) through the funds provided to the Sino-German Collaborative
Research Center TRR110 ``Symmetries and the Emergence of Structure in QCD"
(NSFC Grant No. 12070131001, DFG Project-ID 196253076-TRR 110). 
\end{acknowledgments}



\end{document}